\documentclass[manuscript]{aastex}

\slugcomment{The Astronomical Journal}

\shorttitle{Active Asteroid 324P}
\shortauthors{Jewitt et al.}

\begin{document}

\title{Hubble Space Telescope Observations of \\Active Asteroid 324P/La Sagra}


\author{David Jewitt$^{1,2}$,  Jessica Agarwal$^3$, Harold Weaver$^4$,  Max Mutchler$^5$, Jing Li$^{1}$, \\and Stephen Larson$^6$
}

\affil{$^1$Department of Earth, Planetary and Space Sciences,
UCLA, 
595 Charles Young Drive East, 
Los Angeles, CA 90095-1567\\
$^2$Dept.~of Physics and Astronomy,
University of California at Los Angeles, \\
430 Portola Plaza, Box 951547,
Los Angeles, CA 90095-1547\\
$^3$ Max Planck Institute for Solar System Research, Justus-von-Liebig-Weg 3, 37077 G\"ottingen, Germany \\
$^4$ The Johns Hopkins University Applied Physics Laboratory, 11100 Johns Hopkins Road, Laurel, Maryland 20723  \\
$^5$ Space Telescope Science Institute, 3700 San Martin Drive, Baltimore, MD 21218 \\
$^6$ Lunar and Planetary Laboratory, University of Arizona, 1629 E. University Blvd.
Tucson AZ 85721-0092 \\
}

\email{jewitt@ucla.edu}


\begin{abstract}
Hubble Space Telescope observations of active asteroid 324P/La Sagra near perihelion  show continued mass loss consistent with the sublimation of near-surface ice.  Isophotes of the coma measured from a vantage point below the orbital plane are best matched by steady emission of particles having a nominal size $a \sim$ 100 $\mu$m.  The inferred rate of mass loss, $dM_d/dt \sim$0.2 kg s$^{-1}$, can be supplied by sublimation of water ice in thermal equilibrium with sunlight from an area as small as 930 m$^2$, corresponding to about 0.2\% of the nucleus surface.  Observations taken from a vantage point only 0.6\degr~from the orbital plane of 324P set a limit to the velocity of ejection of dust in the direction perpendicular to the plane, $V_{\perp} <$ 1 m s$^{-1}$.   Short-term photometric variations of the near-nucleus region, if related to rotation of the underlying nucleus, rule out periods $\le$ 3.8 hr and suggest that rotation probably does not play a central role in driving the observed mass loss. We estimate that, in the previous orbit, 324P lost about 4$\times$10$^7$ kg in dust particles, corresponding to 6$\times$10$^{-5}$ of the mass of a 550 m spherical nucleus of assumed density $\rho$ = 1000 kg m$^{-3}$.  If continued, mass loss at this rate would limit the lifetime of 324P to $\sim$1.6$\times$10$^4$ orbits (about 10$^5$ yr). To survive for the 100 Myr to 400 Myr timescales corresponding, respectively, to dynamical and collisional stability requires a duty cycle $2\times 10^{-4} \le f_d \le 8\times 10^{-4}$.  Unless its time in orbit is  over-estimated by many orders of magnitude,  324P is revealed as a briefly-active member of a vast population of otherwise dormant ice-containing asteroids.

\end{abstract}


\section{Introduction}

The active asteroids are solar system bodies which have asteroid-like orbits but which also show  transient,  comet-like activity.  Evidence amassed over the past decade shows that this activity results from a remarkably broad range of physical processes, including impact, thermal fracture, rotational instabilities and ice sublimation.  Identification of these processes in the main-belt represents a scientific watershed, by revealing diverse asteroid  processes that were previously unobserved (Hsieh and Jewitt 2006, Jewitt 2012, Jewitt et al.~2015).

324P/La Sagra (formerly 2010 R2 and hereafter ``324P'') is one of four active asteroids for which mass-loss has been reported on different orbits (the others are 133P/Elst-Pizarro, 238P/Read and 313P/Gibbs; Jewitt et al.~2015).  Mass loss near successive perihelia is a natural indicator of a thermal process, presumably the sublimation of near-surface ice.  Dynamical models show that the capture of ice-rich Jupiter family comets from the Kuiper belt, while possible, is highly inefficient given the current architecture of the solar system (Fernandez et al.~2002, Levison et al.~2006, Hsieh and Haghighipour 2016). Therefore,  it is likely that at least some of the icy objects in the main-belt are primordial residents.  Moreover,  the observed rates of mass loss cannot be sustained for billion-year timescales given the small sizes of most active asteroids. It is therefore likely that these objects spend most of the time between triggering events (impacts?) in an inactive state and that the currently active asteroids represent a much larger population of ice-containing but inert bodies in the asteroid belt.  They are the tip of the iceberg.  While no spectroscopic detection of the gaseous products of sublimation has yet been reported (except in 1 Ceres, Kuppers et al.~2014), the limits to gas production ($\lesssim$1 kg s$^{-1}$) are consistent with the rates of activity estimated from dust.

Active asteroid 324P has semimajor axis, $a$ =  3.096 AU, eccentricity $e$ = 0.154, and inclination $i$ = 21.4\degr, leading to an asteroid-like Tisserand parameter measured relative to Jupiter, $T_J$ = 3.099.  The object was discovered in an active state on UT  2010 September 14.9 (Nomen et al.~2010), three months after perihelion on UT 2010 June 26.  Revived activity was again reported in the summer of 2015 (Hsieh and Sheppard 2015), just prior to the next perihelion on UT 2015 November 30.  The development of  activity in the discovery epoch was described independently in papers by Moreno et al.~(2011) and Hsieh et al.~(2012), while measurements from 2013 (when the object was in an apparently inactive state) were described by Hsieh (2014).  Thermal infrared detections were reported by Bauer et al.~(2012).
In this paper we report new observations from the Hubble Space Telescope taken to investigate 324P  at the highest available angular resolution.

\section{Observations}
Observations  using the Hubble Space Telescope were secured under target-of-opportunity program GO 14263 (two orbits) and continued with so-called mid-cycle time under GO 14458 (four orbits).  A journal of observations is provided in Table (\ref{geometry}).  The earliest observations (UT 2015 September 28) were obtained from a vantage point 9.5\degr~below the orbital plane of 324P, affording a view of the distribution of dust particles across the orbital plane.  The last observations, on UT 2015 December 18, were scheduled to coincide with the passage of the Earth through the orbital plane (in practice, the out-of-plane angle was -0.6\degr).  From this nearly in-plane geometry, the images provide, instead, a complementary measure of the distribution of dust in a direction perpendicular to the orbit plane.  All observations  employed the broadband F350LP filter, which has effective wavelength 6230\AA~and full-width-at-half-maximum (FWHM) of 4758\AA~when observing a sun-like spectrum.  This filter provides maximum sensitivity to faint and low surface brightness targets, and is ideally suited to the observation of weak cometary activity.    The images have a scale of 0.04\arcsec~pixel$^{-1}$, corresponding to 65 km pixel$^{-1}$ and 90 km pixel$^{-1}$ on the first and last dates of observation, respectively. Figure (\ref{composite}) shows, for each date of observation, the median of five, 420 s integration images taken within a single orbit,  along with arrows to mark  the projected anti-solar direction (yellow) and the negative heliocentric velocity vector (green).

\subsection{Dust Distribution}
\begin{figure}
\plotone{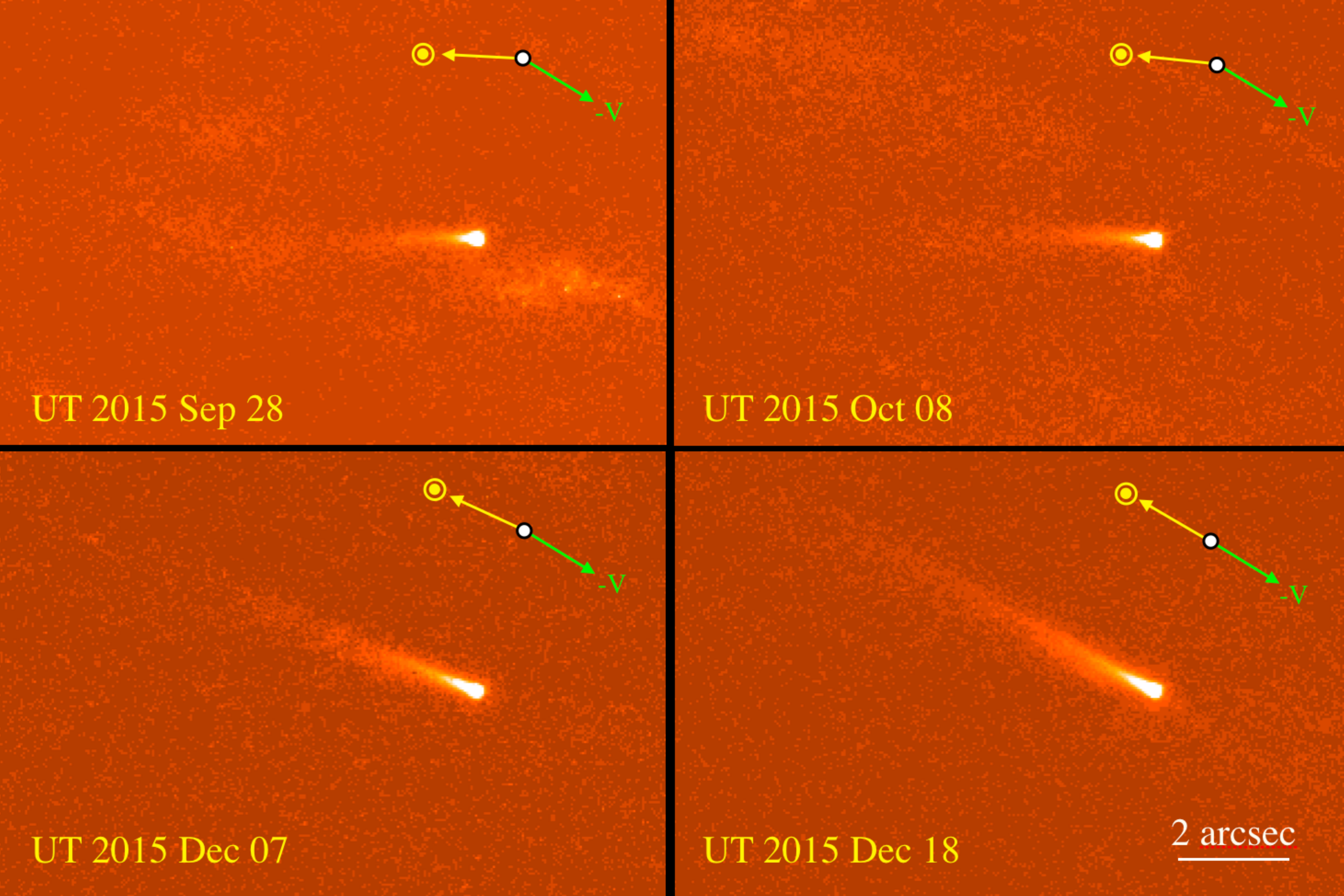}
\caption{Images of 324P at the four epochs of HST observation.  The panels have North to the top, East to the left and are shown at the same angular scale, indicated by the 2\arcsec~scale bar in the lower right.   Yellow arrows indicate the projected antisolar direction while green arrows show the negative of the projected heliocentric velocity vector, for each epoch.  Faint, imperfectly removed background objects are apparent, particularly in the September 28 observation.  \label{composite}}
\end{figure}

Inspection of Figure (\ref{composite}) shows that the tail  of 324P  maintains a position angle that  closely follows the changing antisolar direction.  The position angle of the projected anti-solar vector changes considerably (from $\sim$87\degr~to $\sim$60\degr) between the September and December observations, while the position angle of the projected orbit remains relatively constant near 238\degr~(Table \ref{geometry}).   The swing of the tail is as expected for recently-released particles that are small enough to be strongly accelerated by solar radiation pressure.  In contrast, we observe in Figure (\ref{composite}) no clear indication for the presence of dust particles along the direction of the projected orbit, suggesting that the abundance of large, radiation pressure-insensitive particles is small.  

To explore these qualitative inferences further, we computed the trajectories followed by dust particles ejected from the nucleus of 324P.  The latter trajectories are functions of the dust ejection velocity, $v$, and of $\beta$, the ratio of the radiation pressure induced acceleration to the local solar gravitational acceleration, $g_{\odot}$.  Ratio $\beta$ depends on the particle morphology (e.g.~compact sphere vs.~fluffy aggregate) and on the composition (Bohren and Huffman 1983). For compact dielectric spheres, $\beta$ is  proportional to the inverse particle radius (provided $\beta <$ 1) and, to a useful level of approximation, we write $\beta \sim a_{\mu m}^{-1}$, where $a_{\mu m}$ is the particle radius expressed in microns.  

\begin{figure}
\epsscale{.950}
\center
 \includegraphics[angle=270,height=8.0cm,origin=l]{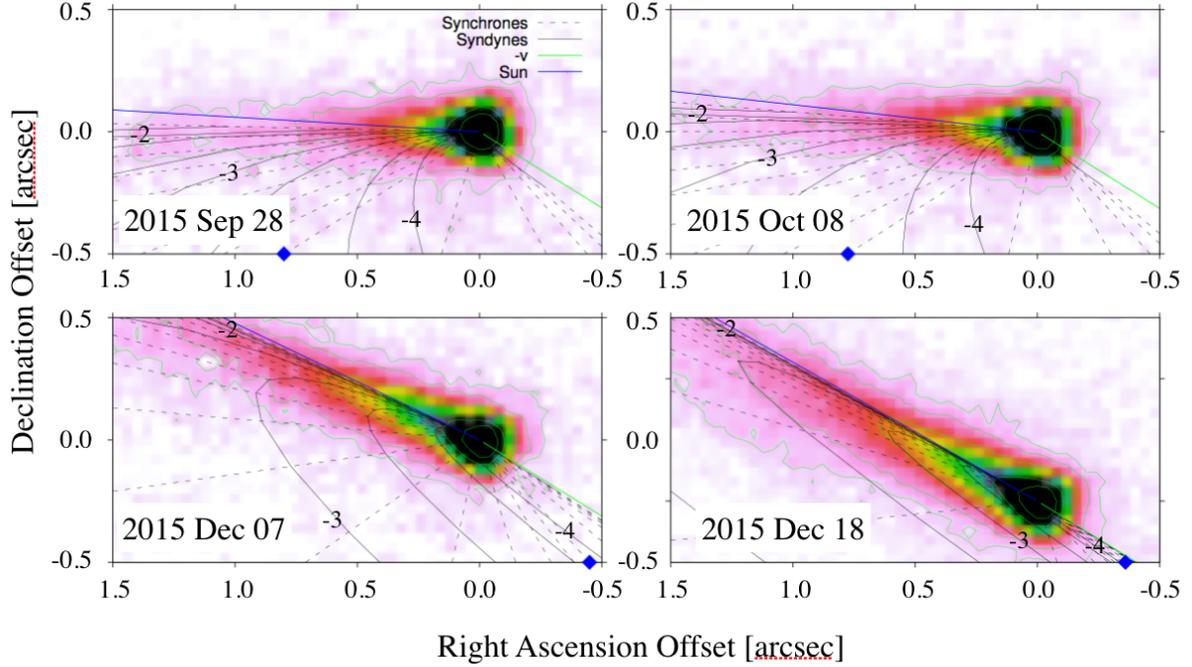}
 \endcenter
\caption{Syndyne-synchrone models computed as described in the text.  Syndynes (solid black lines) correspond to particles with $\beta$ = 2$\times$10$^{-2}$, 1$\times$10$^{-2}$, 5$\times$10$^{-3}$, 2$\times$10$^{-3}$, 1$\times$10$^{-3}$, 5$\times$10$^{-4}$, 2$\times$10$^{-4}$, 1$\times$10$^{-4}$ and 1$\times$10$^{-6}$ measured counterclockwise about the nucleus.  For clarity, we label only the syndynes having $\log_{10}\beta$ = -2, -3 and -4.  Synchrones are shown as straight dashed lines, with the synchrone of 2015 July 01 marked with a blue diamond. Synchrones clockwise from this date (measured with the nucleus as center) are plotted  at 10-day intervals, while those counter-clockwise from 2015 July 01 have 100-day intervals.  Solid blue and green lines mark the projected anti-solar and negative orbital velocity vectors, respectively.
\label{synsyns}}
\end{figure}

Solid lines in Figure (\ref{synsyns})  show syndynes, which are the loci of positions of particles of a given $\beta$ released from the nucleus with negligible velocity at different times (c.f. Finson and Probstein 1968).  Dashed lines in the figure show synchrones, which mark the loci of positions of particles having different $\beta$ but which are all released from the nucleus at a given time.  The direction and slight counter-clockwise curvature of the tail are best matched by syndynes with $\beta \sim$ 0.005 to 0.02, corresponding to particle radii $a \sim$ 50 to 200 $\mu$m.  We take $a \sim$ 100 $\mu$m as the nominal grain size in 324P.  While evidence for larger particles (smaller $\beta$) is distinctly absent, we note that such particles are always difficult to detect in comets. This is because they  contain a small fraction of the scattering cross-section and because their small acceleration by solar radiation pressure may limit their abundance far from the nucleus.

\subsection{Photometry}
Evidently, 324P is an extended source. Photometric measurements employing  fixed-angle apertures sample a larger volume of coma  at larger geocentric distances, potentially making the object appear artificially bright as it receeds from Earth.  To examine the magnitude of this ``aperture effect'', we determined apparent magnitudes in two ways.   First, in Table (\ref{photometry}) we list measurements within circular apertures of fixed angular radii 0.2\arcsec, 1.0\arcsec, 4.0\arcsec~and 6.0\arcsec.  The background for these measurements was determined from the median signal within a concentric annulus having inner and outer radii 6.0\arcsec~and 12.0\arcsec, respectively.  Second,  in Table (\ref{absolute}) we list measurements within circular apertures having fixed linear radii (projected to the distance of the object) of 460 km, 2300 km, 9200 km and 13,800 km.  The fixed linear radius measurements employed background subtraction from a concentric annulus extending from 13,800 km to 27,600 km.  In both Tables (\ref{photometry}) and (\ref{absolute}), the photometric errors increase  with aperture radius  because of the growing importance of sky background uncertainties in the larger aperture measurements.   Comparison of Tables (\ref{photometry}) and (\ref{absolute}) shows that the aperture effect is small, because 324P is very centrally-condensed. Nevertheless, we use only the fixed linear-aperture measurements from Table (\ref{absolute}) in the following discussion. 

The apparent magnitudes, $V$,  were converted to absolute values by correcting to unit heliocentric, $R$, and geocentric, $\Delta$, distance and to zero phase angle, $\alpha$, using the inverse square law

\begin{equation}
H_{V} = V - 5\log(R\Delta) + 2.5\log \Phi(\alpha).
\label{absolute}
\end{equation}

\noindent Here, $0 \leq \Phi(\alpha) \leq 1$ is the phase function, equal to the ratio of the scattered light at phase angle $\alpha$ to that at $\alpha$ = 0\degr.   We assumed the phase function formalism of Bowell et al.~(1989) with parameter $g$ = 0.15, as appropriate for a C-type object and $g$ = 0.25, for an S-type spectrum.  The phase function of 324P is unmeasured, introducing  uncertainty into the value of $H_V$ over and above that due to measurement errors.   At the largest phase angles of the present observations ($\alpha$ = 22\degr, Table \ref{geometry}), the difference between assumed C-type and S-type phase corrections is $\sim$0.1 magnitudes, giving an estimate of the magnitude of the phase correction uncertainty.   Absolute magnitudes using $H_V(C)$ are given in Tables (\ref{photometry}) and Table (\ref{absolute}) together with their statistical uncertainties.

The absolute magnitudes are related to the effective scattering cross-section of the material within the photometry aperture, $C_e$ (km$^2$),  by

\begin{equation}
C_e = \frac{2.24\times10^{16} \pi}{p_{\lambda}} ~10^{0.4[m_{\odot, \lambda} - m_{\lambda}(1,1,0)]}
\label{area}
\end{equation}

\noindent where $p_{\lambda}$ is the geometric albedo of 324P and $m_{\odot, \lambda}$ is the apparent magnitude of the Sun, both at wavelength $\lambda$.  We assume $V_{\odot}$ = -26.77.  Bauer et al.~(2012) obtained a 3$\sigma$ upper limit to the visual geometric albedo $p_V \le$ 0.06; we assume $p_V$ = 0.05.  The resulting scattering cross-sections are listed in Table (\ref{photometry}), computed assuming $p_{V}$ = 0.05 with small adjustments for $p_B$ and $p_R$ as indicated by the broadband colors.  Uncertainties on $C_e$  reflect the larger ($\sim 0.1$ magnitude) systematic uncertainties estimated from the difference between the C-type and S-type phase functions.

\begin{figure}
\plotone{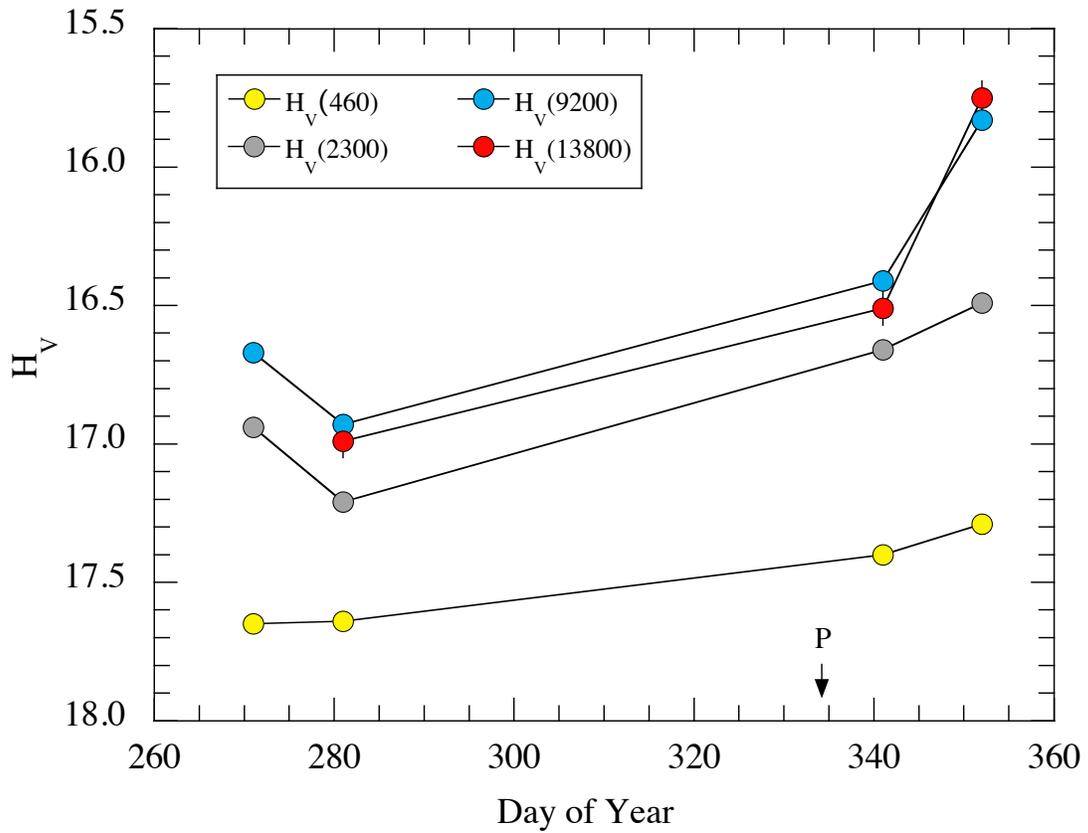}
\caption{Absolute magnitude as a function of time (representated as Day of Year in 2015, with 1 = UT 2015 January 1) and of aperture size.   Brightening of the object is evident in the later data. P marks the date of perihelion. \label{HVplot}}
\end{figure}

Figure (\ref{HVplot}) shows the absolute magnitudes, $H_V$, as a function of time, measured as day-of-year in 2015, for four apertures of fixed linear radius.  The absolute magnitudes generally brighten with time, showing that material is being actively released from the nucleus into the dust tail over a $\sim$2 month interval around perihelion (day of year = 334).  The largest (13,800 km) aperture photometry brightens from  $H_V$ = 17.00 to 15.75, a factor of 3 in cross-section (Table \ref{absolute}).  The brightening corresponds to the release of a mass of dust

\begin{equation}
M_d = \frac{4}{3} \rho \overline{a} \Delta C_e
\end{equation}

\noindent where $\rho$ is the mass density of the dust grains, $\overline{a}$ is their mean radius and $\Delta C_e$ is the change in the scattering cross-section.  We take $\overline{a}$ = 100 $\mu$m,  as suggested by the syndynes in Figure (\ref{synsyns}),  $\Delta C_e$ = 9.4$\pm$1.5 km$^2$ between October 08 and December 18 from Table (\ref{absolute}) and assume $\rho$ = 1000 kg m$^{-3}$ to obtain a mass ejection $M_d$ = 1.3$\pm$0.2$\times$10$^6$ kg.  If ejected steadily over this 71 day interval, the average dust ejection rate would be $dM_d/dt \sim$ 0.2 kg s$^{-1}$.  

We solved the energy balance equation for an exposed, perfectly absorbing water ice surface located at the sub-solar point on 324P. At $r_H$ = 2.62 AU, we find that ice would sublimate, in equilibrium with sunlight, at the specific rate $F_s$ = 4.3$\times$10$^{-5}$ kg m$^{-2}$ s$^{-1}$.  The area of exposed ice needed to supply dust at the rate $dM_d/dt$ is given by 

\begin{equation}
A_s = \frac{dM_d/dt}{f_{dg} F_s}
\label{subl_area}
\end{equation}

\noindent where $f_{dg}$ is the ratio of the dust to gas mass production rates.  Measurements of Jupiter family comets generally show $f_{dg} > 1$.  For example, a detailed investigation of 67P/Churyumov-Gerasimenko by Fulle et al.~(2016) gave $f_{dg} \sim$ 5 to 10 while values 10 $\le f_{dg} \le$ 30 were obtained in comet 2P/Encke by Reach et al.~(2000).  We conservatively adopt $f_{dg}$ = 5 to find $A_s$ = 930 m$^2$ (only $\sim$0.02\% of the surface of a spherical nucleus of radius 550 m) corresponding to a circular patch as small as $r_s = (A_s/\pi)^{1/2}$ $\sim$17 m in radius.  This is a lower limit to $A_s$ in the sense that we have assumed sublimation at the maximum possible rate by placing the ice patch at the subsolar point and assuming that it is perfectly absorbing.  Ice that is reflective, or located away from the subsolar point or buried beneath a thin, insulating dust mantle would sublimate less rapidly and require a larger area to supply the dust mass loss rate.   Indeed, the much brighter coma observed in 2010 would require a larger area of exposed ice, $A_s \gtrsim$ 19,000 m$^2$ (about 0.5\% of the surface area, or $r_s \gtrsim$ 80 m), if produced by sublimation in steady-state. In either case, it is evident that sublimation from a very small fraction of the nucleus surface can supply the dust release in 324P near perihelion.   We checked to confirm that the drag from a gas flow of strength $F_s$ is more than sufficient to launch spherical particles of 100 $\mu$m size against the gravity of a nucleus estimated at only 0.55 km in radius (Hsieh 2014).  Indeed, neglecting adhesive contact forces, particles up to $\sim$0.5 m could be launched by gas drag.

\subsection{Dust Profiles}
\begin{figure}
\plotone{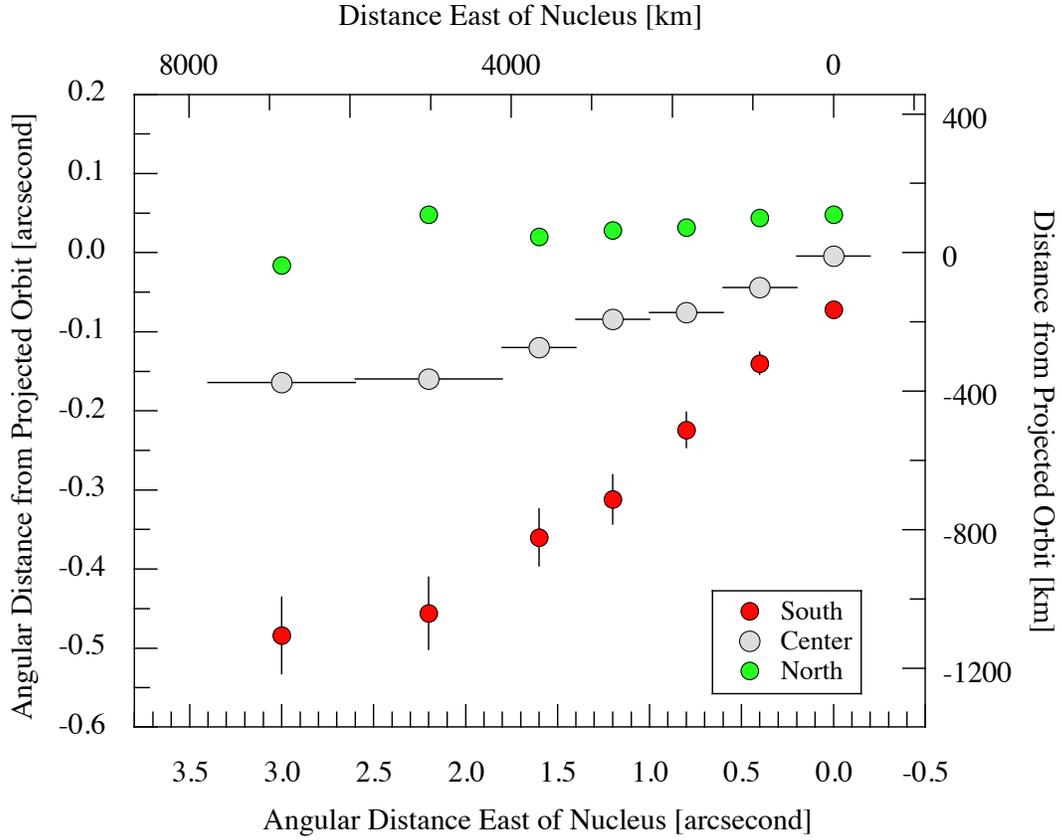}
\caption{Distance from projected orbit measured as a function of separation measured east from the nucleus.  The locations of peak brightness and of  half-peak brightness are measured separately to the north and the south of the trail.  Horizontal bars show the range of distances within which the faint tail signal was binned while (vertical) error bars are estimates of the measurement uncertainty as described in the text.  \label{vertprofile}}
\end{figure}

We use the surface brightness profile of the dust in the direction perpendicular to the projected orbit plane to set a constraint on the dust ejection velocity.    For this purpose, we rotated the composite image taken on  UT 2015 December 18 about the nucleus in order to bring the projected orbit to the horizontal, and then computed  vertical profiles averaged along segments of the tail.  Results from the profiles are shown in Figure (\ref{vertprofile}), where three quantities are plotted.  Gray circles show the location of peak dust brightness together with horizontal bars to indicate the width of each segment of the tail used to average the data.  The distances to the half-power points of each profile to the North and South sides of the tail are shown as green and red circles, respectively.  Error bars on the latter grow with distance from the nucleus because of the growing effect  of sky noise and background sky brightness errors (caused by contamination from faint, trailed background sources) as the tail surface brightness rapidly declines.    

In all cases, the dust is more extended to the south of the tail axis than to the north.  Figure (\ref{synsyns}) shows that this is a projection effect, caused by viewing the dust sheet from a vantage point slightly below the plane (by -0.6\degr; c.f.~Table \ref{geometry}).  We take the extension of the dust to the north of the axis as the most meaningful measure of the out-of-plane distribution.  This extension is about 0.05\arcsec~(linear distance $h_{w} \sim$100 km) at 2\arcsec~($\ell \sim$4500 km) from the nucleus, far below the typical resolution afforded by ground-based telescopes.  

Dust motion perpendicular to the orbit plane is unaffected by radiation pressure on timescales short compared to the orbital period; the perpendicular distance travelled in time $t$ is just $h_{w} = v_{\perp} t$, where $v_{\perp}$ is the perpendicular velocity.  Dust motion parallel to the orbit is accelerated by radiation pressure, such that the distance travelled after time $t$ is $\ell = \beta g_{\odot}(1) t^2/2 r_H^2$. Here, $g_{\odot}(1)$ = 0.006 m s$^{-2}$ is the gravitational attraction to the Sun at $r_H$ = 1 AU, and $r_H$ is the heliocentric distance expressed in AU.  Eliminating $t$ gives 

\begin{equation}
V_{\perp} = \left(\frac{\beta g_{\odot}(1)}{2 r_H^2 \ell}\right)^{1/2} h_{w}.
\label{speed}
\end{equation}

\noindent Calculations (Silsbee and Draine 2016) show that silicate particles have $\beta <$ 1, regardless of their size or aggregate structure (small, metal particles can have $\beta \ge$ 1  but these are of dubious relevance in the active asteroid population).   With  $\beta <$ 1, $r_H$ = 2.621 AU and $h_w \sim$ = 100 km at $\ell$ = 4500 km, Equation (\ref{speed}) gives $V_{\perp} <$ 1 m s$^{-1}$, which we take as an upper limit to the dust ejection velocity.  

In the classical comet model (Whipple 1950) the terminal dust velocity resulting from gas drag varies approximately as $v \propto \beta^{1/2}$. Micron-sized grains are dynamically well-coupled to the gas, with terminal speeds near the sound speed (which we take as $v_s \sim$ 450 m s$^{-1}$ at $r_H$ = 2.621 AU).  We therefore expect that, in the classical model, the optically dominant $\sim$100 $\mu$m sized particles should have $v \sim$ 45 m s$^{-1}$, considerably larger than the upper limit to $V_{\perp}$ set by the width of the dust tail using Equation (\ref{speed}).  Similarly  low dust speeds have been reported in other active asteroids (Jewitt et al.~2014) and explained as a consequence of sublimation from a small area source.  A small sublimating area limits the path length over which expanding gas can accelerate entrained dust and so leads to lower terminal velocities.  From  a source of horizontal dimension $r_s$, the terminal velocity is given by Equation (A5) of Jewitt et al.~(2014) as

\begin{equation}
V_T = \left(\frac{3 C_D V_g F_s(r_n) r_s}{4 \rho a} - \frac{8\pi G \rho_n r_n^2}{3}\right)^{1/2}.
\label{VT}
\end{equation}

\noindent Here, $C_D \sim 1$ is a dimensionless drag coefficient, $G$ is the gravitational constant, $\rho$ is the density of the ejected grain, $\rho_n$ is the density of the nucleus, $r_n$ is its radius and the other symbols are as defined above.  We assume $\rho$ = $\rho_n$ = 1000 kg m$^{-3}$ and take $a$ = 100 $\mu$m, $r_s$ = 80 m to find $V_T \sim$ 3 m s$^{-1}$. This is an order of magnitude smaller than the speed estimated from $v \propto \beta^{1/2}$ scaling and more comparable to, but slightly larger than, the measured upper limit $V_{\perp} <$ 1 m s$^{-1}$.  With our measurements we cannot reject the possibility that the dust is ejected slowly by a process other than gas drag, but such an alternative explanation would struggle to account for the recurrence of activity at two successive perihelia.

\subsection{Nucleus Photometry and Rotation}
Photometry of the nucleus was obtained using small apertures fixed in linear radius (460 km) projected to the distance of 324P (Table \ref{absolute}).  The faintest absolute magnitude, $H_V$ = 17.65$\pm$0.01 on UT 2015 September 28, is brighter than reported in data from the previous orbit.  Hsieh (2014) found absolute red magnitude $H_R$ = 18.4$\pm$0.2 to 18.7$\pm$0.2 (depending on the adopted phase angle correction) which, assuming a sun-like color index V-R = 0.35 (Holmberg et al.~2006), corresponds to $H_V$ = 18.7$\pm$0.2 to 19.0$\pm$0.2.  The nucleus on UT September 28 is thus brighter by a magnitude or more, presumably indicating ongoing dust release in the recent data.  For this reason, Hsieh's (2014) determination of the effective radius of the nucleus, $r_n$ = 0.55$\pm$0.05 km (assumed geometric albedo 0.05), remains the most trustworthy.

Three contiguous HST orbits on UT 2015 December 18 (Table \ref{geometry}) were secured specifically to provide a  timebase  sufficient to assess short-term variations in the scattered light.  To measure these images, we removed cosmic ray artifacts by hand using the following procedure.  First, we computed the median image from the 5 images taken within each orbit of HST.  The use of the median effectively eliminates cosmic rays that are abundant in the individual images.   Next, we subtracted this median image from each of the individual images in the coresponding orbit so as to remove the ``steady'' signals in the data, leaving only cosmic rays, noise and the residuals of passing background objects.  Near-nucleus cosmic rays were then removed by interpolation of the brightness in surrounding pixels.  The final step was to add back the median image from each orbit, to produce a set of cosmic ray cleaned data suitable for nucleus photometry.  This procedure worked well in all but a few images for which cosmic ray and trailed field galaxy contamination was too severe to be removed.  We rejected such images from further consideration.

The nucleus photometry was measured using projected circular apertures 0.2\arcsec~in radius.  The background  was determined using the median signal computed within a surrounding annulus having inner and outer radii 0.2\arcsec~and 0.4\arcsec, respectively.  This background annulus includes a contribution from the dust component of 324P but a comparison of measurements in Table (\ref{nucleus_phot}) with photometry obtained using a larger sky annulus (Table \ref{photometry}) shows that the dust has only a small influence on the photometry.   

\begin{figure}
\plotone{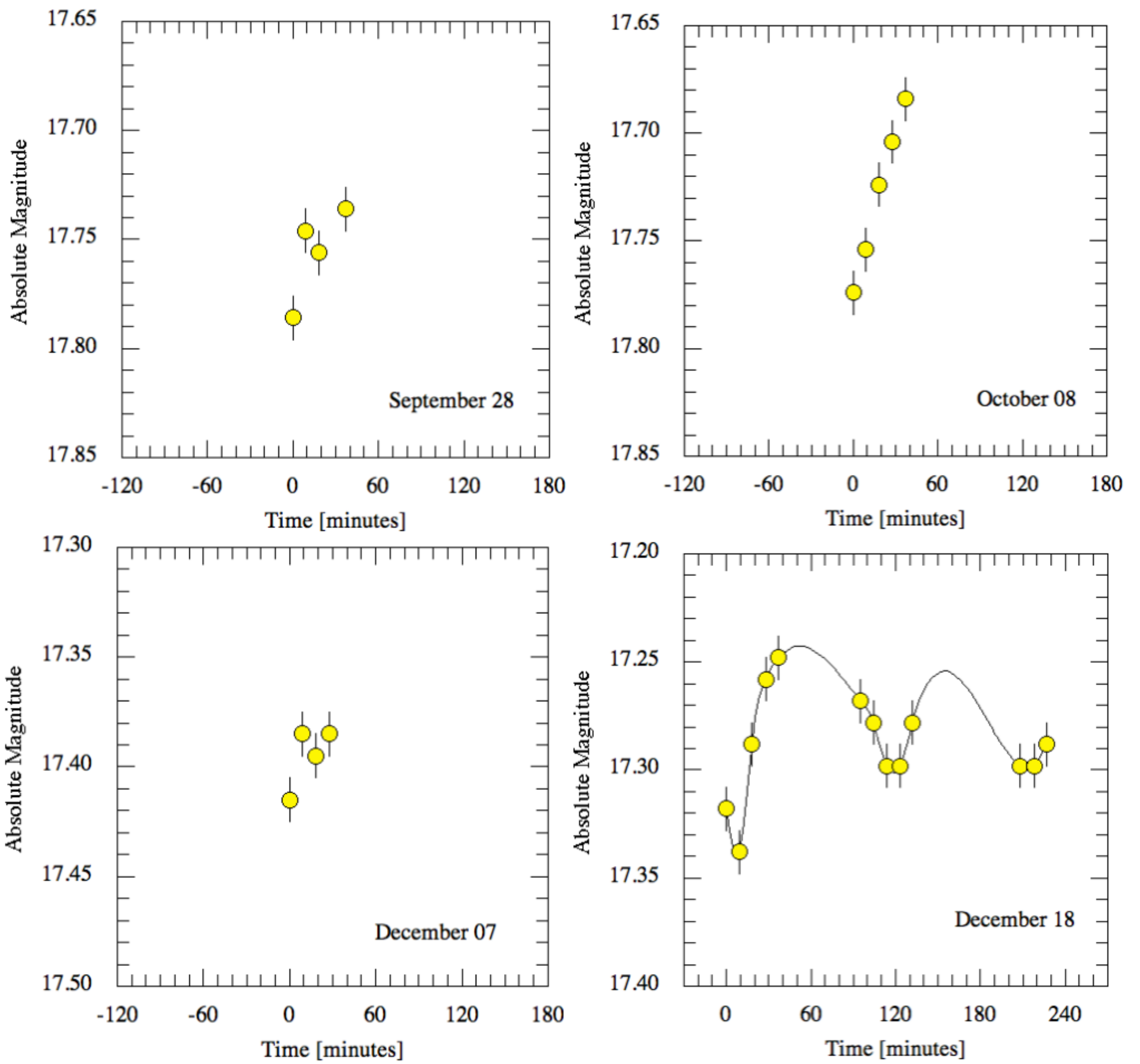}
\caption{Lightcurves of 324P at the four epochs of HST observation. The times are given in minutes elapsed from the first observations listed in Table (\ref{geometry}). A cubic spline fit to the data is plotted in the panel for UT 2015 December 18 to guide the eye.  \label{lightcurves}}
\end{figure}

The results from Table (\ref{nucleus_phot}) are plotted in  Figure (\ref{lightcurves}), where brightness variations up to $\sim$0.1 magnitude are seen to exist at each epoch of observation.  The variations are large compared to the uncertainties of measurement, which we estimate to be $\pm$0.01 magnitudes, and thus must be considered real.  The December 18 data have the longest timebase and clearly show   brightness variations on timescales that are consistent with the effects of nucleus rotation, while single-orbit visits on September 28, October 08 and December 07 provide only brief snapshots. It is impossible to uniquely link the photometric variations across all four dates of observation because of the long intervals between observations and the resulting aliasing in the data.  However, we note that the rising brightness recorded on UT 2015 October 08 (upper right panel of Figure \ref{lightcurves}) very likely corresponds to the rising section of the lightcurve on December 18 (lower right panel), meaning that the rotation period must be a sub-multiple of the interval between these dates.  

Given the nucleus absolute magnitude, $H_V$  = 18.7$\pm$0.2 as described above, is so much fainter than the values plotted in Figure (\ref{lightcurves}) it is obvious that dust must contribute to the total cross-section.  For example, on UT 2015 September 28 about half the cross-section in the 460 km radius aperture (c.f.~Table \ref{absolute}) can be attributed to a 0.55 km radius spherical nucleus while, by UT 2015 December 18, this fraction had fallen to $\sim$1/3rd.  However, it is unlikely that near nucleus dust, even though it dominates the cross-section,  is responsible for the observed photometric variability.  To see this we note that dust traveling at the maximum speed allowed by the tail width, namely $\sim$1 m s$^{-1}$ would take 4.5$\times$10$^5$ s ($\sim$5 day) to cross the 460 km radius aperture and, therefore, would contribute a quasi-constant background  to the daily photometry.  In particular, the decrease in brightness between $\sim$60 minutes and $\sim$120 minutes in the December 18 panel of the Figure cannot be explained by dust escaping the photometry aperture unless the dust speed is $\geq$130 m s$^{-1}$, which is inconsistent with the dust tail width.  The variability in the Figure is more naturally interpreted as a nucleus rotational lightcurve diluted by near-nucleus coma.  The $\sim$0.1 magnitude range of variation is a lower limit to the actual range of the rotating bare nucleus because of the adjacent dust.  

While the sampling of the HST data is too limited to be able to define the nucleus rotation period, we can nevertheless use the data to place physically important limits on the rotation.  Crucially, the lightcurve on December 18 does not repeat, showing that the rotation period must be longer than the 3.8 hr interval between the first and last images in this sequence.  Thus, we are confident that the rotation period of the nucleus of 324P is longer than the critical 2.2 hour  ``rotational barrier'' period shown by suspected rubble-pile asteroids (Harris 1996, Pravec et al.~2002).  In turn, this provides no reason to think  that mass loss from 324P might be strongly influenced by rotational instability, unless the nucleus is a highly-elongated, prolate body or one having extremely low density. Future time-resolved photometry of 324P when in an inactive state is needed to establish the rotational state. 

\section{Discussion}
\begin{figure}
\plotone{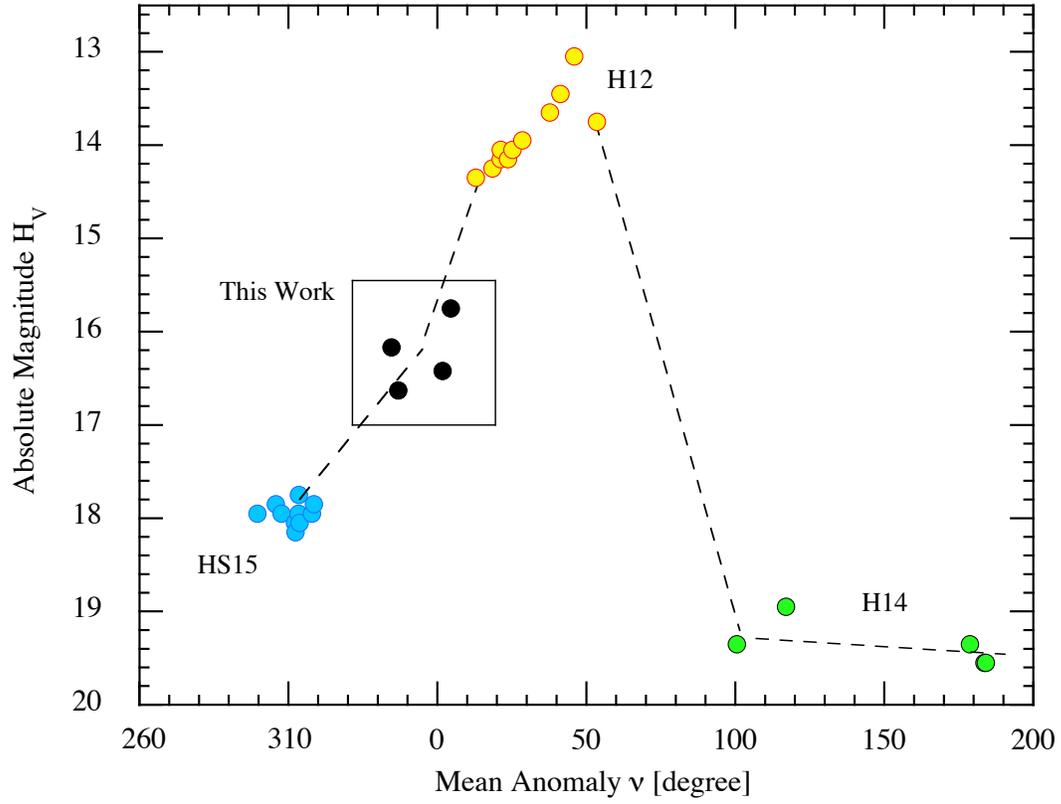}
\caption{Reported absolute magnitudes as a function of the mean anomaly.  We combine data from Hsieh et al.~(2012): H12, Hsieh (2014): H14, and Hsieh and Sheppard (2015): HS15 with the measurements from the present paper.     \label{mags}}
\end{figure}

The present absolute magnitudes are compared with published measurements as a function of the mean anomaly, $\nu$, in Figure (\ref{mags}).  In making this plot, all photometry has been corrected to the V filter assuming V-R = 0.35.  The Figure shows that our measurements are intermediate between the weakly active state recorded earlier in 2015 (Hsieh and Sheppard 2015) and the bright state in which 324P was discovered in 2010 (Hsieh et al.~(2012) and gives the impression that the absolute brightness is a nearly monotonic function of the mean anomaly angle. The persistence of activity from pre-perihelion $\nu \sim$ 310\degr~to post-perihelion $\nu \sim$ 100\degr~corresponds to a $\sim$660 day interval during which the heliocentric distance changes from 2.7 AU inbound to 3.1 AU outbound.    However, whether this is a real, temporal progression that repeats from orbit to orbit cannot be known until more observations are acquired.     Unfortunately, the elongation of 324P fell below 50\degr~immediately following the observations reported here, terminating measurements from HST.   Observations over the next year will be of great interest in showing whether or not the photometric behavior of 324P is repeatable from orbit to orbit.

The dust properties in the 2010 active phase were modeled by Moreno et al.~(2011).  They assumed $v \propto a^{-1/2}$ and fitted the isophotes of the dust to find a grain speed $v$ = 0.2 m s$^{-1}$ for $a$ = 10$^{-2}$ m, with an uncertainty of order a factor of two.  Scaling to 100 $\mu$m particle size would give $v \sim$ 2 m s$^{-1}$, compared with our upper limit $V_{\perp} <$ 1 m s$^{-1}$.  The mass loss rates estimated by Moreno et al.~peaked near 4 kg s$^{-1}$, which is larger by about a factor of 20 than the $\sim$0.2 kg s$^{-1}$ deduced from our photometry in 2015.  However, 324P was intrinsically much brighter when observed by Moreno et al.   For example, Figure (\ref{mags}) shows that the average absolute magnitude in our data ($H_V \sim$ 16.4) is fainter than the peak brightness recorded in 2010 ($H_V \sim$ 13.0) by 3.4 magnitudes, corresponding to a factor $\sim$23.  Using similar arguments, Hsieh and Sheppard (2015) found $dM_d/dt \sim$ 0.1 kg s$^{-1}$ near $\nu$ = 300\degr.

We estimated the total mass loss from 324P as follows.  We convert the instantaneous absolute magnitude to production rate using $dM_d/dt = 6\times 10^5~10^{-H_V/2.5}$ and then integrate with respect to time.  From observations in the discovery epoch (i.e.~mean anomalies 12\degr~$ \le \nu \le$ 54\degr, shown as yellow circles in Figure \ref{mags}) we obtain $\Delta M_d$ = 2$\times 10^7$ kg.  Integrating over the full range of observations (280 $\le \nu \le$ 100\degr) gives $\Delta M_d \sim$ 4 $\times 10^7$ kg, which provides an order of magnitude estimate of the mass lost from the nucleus in one orbit.   The mass of a 550 m radius sphere having the  density $\rho$ = 10$^3$ kg m$^{-3}$ is $M_n =6 \times$10$^{11}$ kg.    If continued, mass loss would deplete all the mass in the nucleus of 324P in a time $t \sim (M_n/\Delta M_d) t_K$, where $t_K \sim$ 5 yr is the average Keplerian orbital period. Substituting, we find $t \sim$ 10$^5$ yr.  

To survive for time, $T$, would require a duty cycle (the fraction of  time over which the body is active) given by $f_d \sim (M_n/\Delta M_d)(t_K/T)$.  Hsieh et al.~(2012) used numerical integrations to show that the orbit of 324P is stable on timescales $T \sim 10^8$ yr, while the collisional lifetime of a $\sim$1 km diameter asteroid is $T \sim$ 4$\times$10$^8$ yr (c.f.~Figure 14 of Bottke et al.~2005). Substituting these timescales we obtain $2\times 10^{-4} \le f_d \le 8\times10^{-4}$.  A somewhat smaller number, $f_d \gtrsim$ 2$\times$10$^{-5}$, was obtained  from statistical arguments based on observations of four active asteroids, summarized in Jewitt et al.~(2015).  

In steady-state, the existence of $n_0$  known,  repetitively active (ice-containing) objects  corresponds to a dormant population, $N = n_0/f_d$.  We set $n_0$ = 4 (active asteroids 133P, 238P, 313P and 324P) and substitute $2\times 10^{-4} \le f_d \le 8\times 10^{-4}$ to find $5\times 10^3 \le N \le 2\times 10^4$.   These are almost certainly under-estimates of the dormant object population, both because we are sensitive only to objects in which the ice is close enough to the physical surface to be occasionally exposed to sunlight (e.g.~Haghighipour et al.~2016) and because the surveys used to identify active asteroids are incomplete (Waszczak et al.~2013, Hsieh et al.~2015).

%

\section{Summary}
We obtained Hubble Space Telescope images of active asteroid 324P on four occasions between UT 2015 September 28 and December 18.  We find that:

\begin{enumerate}
\item High resolution images of 324P show a point-like, but still active, nucleus and a radiation-pressure swept dust tail consisting of $\sim$100 $\mu$m sized particles.  

\item An increase  by a factor $\sim$3 in the near-nucleus dust cross-section  indicates the continuing ejection of dust of mass $M_d \sim$1.3$\pm$0.2$\times$10$^6$ kg.  The average dust production rate in this period, $dM_d/dt \sim$ 0.2 kg s$^{-1}$,  could be supplied by the sublimation of  water ice  covering as little as $A_s \sim$ 930 m$^2$ (about 0.2\%) of the nucleus surface.  The total mass lost in the orbit following the discovery of 324P is estimated at $\sim$4$\times$10$^7$ kg.

\item Observations taken from near the orbital plane of 324P indicate that the dust ejection velocity is $<$1 m s$^{-1}$ (for 100 $\mu$m sized particles).  This low speed is compatible with a model of gas drag acceleration from a sublimating  ice patch that is small compared to the nucleus.  

\item The bare nucleus of 324P was not observed in our data.  However, we detect short-term photometric variations in the near-nucleus region that are likely caused by nucleus rotation.  If so interpreted, these variations  rule-out periods $<$3.8 hr and suggest that rotational instability is unlikely to play a leading role in the loss of mass from this object. 

\item The small size of the nucleus limits the mass-loss  lifetime of 324P to about 10$^5$ yr, which is short compared both to the reported (10$^8$ yr) dynamical lifetime   and to the estimated collisional lifetime (4$\times 10^8$ yr).  Unless the time spent by 324P in orbit is severely over-estimated, the persistence of this object implies a small duty cycle ($2\times 10^{-4} \le f_d \le 8\times 10^{-4}$) and a corresponding  population ($5\times 10^3 \le N \le 2\times 10^4$) of dormant asteroidal counterparts containing near-surface ice.

\end{enumerate}

\acknowledgments

We thank the anonymous referee for a prompt review. Based on observations made with the NASA/ESA Hubble Space Telescope, obtained  at the Space Telescope Science Institute, which is operated by the Association of Universities for Research in Astronomy, Inc., under NASA contract NAS 5-26555. These observations are associated with GO programs 14263 and 14458.



{\it Facilities:}  \facility{HST (WFC3)}.




\clearpage








\clearpage

\begin{deluxetable}{lcccrccccr}
\tablecaption{Observing Geometry 
\label{geometry}}
\tablewidth{0pt}
\tablehead{ \colhead{UT Date and Time} & DOY\tablenotemark{a}   & $\Delta T_p$\tablenotemark{b} & $\nu$\tablenotemark{c} & \colhead{$r_H$\tablenotemark{d}}  & \colhead{$\Delta$\tablenotemark{e}} & \colhead{$\alpha$\tablenotemark{f}}   & \colhead{$\theta_{\odot}$\tablenotemark{g}} &   \colhead{$\theta_{-v}$\tablenotemark{h}}  & \colhead{$\delta_{\oplus}$\tablenotemark{i}}   }
\startdata

2015 Sep 28 23:13 - 23:49 & 271 & -63 & 344.6  	&  2.632 	& 2.213 	& 21.7 	& 86.6 	& 238.1 	& -9.5 \\
2015 Oct 08 09:02 - 09:39 & 281 & -53 & 346.9  	&  2.629 	& 2.326 	& 22.2 	& 83.7 	& 237.9 	& --8.8 \\
2015 Dec 07 03:41 - 04:18 &   341 & 7 &  1.8 & 2.620 & 3.021 & 18.4 & 64.5 & 238.1 & -1.9 \\
2015 Dec 18 00:32 - 04:20 & 352 & 18 & 4.5 & 2.621 & 3.128 & 16.9 & 60.4 & 238.4 & -0.6 \\

\enddata


\tablenotetext{a}{Day of Year, UT 2015 January 01 = 1}
\tablenotetext{b}{Number of days from perihelion (UT 2015-Nov-29.96 = DOY 334). Negative numbers indicate pre-perihelion observations.}
\tablenotetext{c}{True anomaly, in degrees}
\tablenotetext{d}{Heliocentric distance, in AU}
\tablenotetext{e}{Geocentric distance, in AU}
\tablenotetext{f}{Phase angle, in degrees}
\tablenotetext{g}{Position angle of the projected anti-Solar direction, in degrees}
\tablenotetext{h}{Position angle of the projected negative heliocentric velocity vector, in degrees}
\tablenotetext{i}{Angle of Earth above the orbital plane, in degrees}

\end{deluxetable}

\clearpage

\begin{deluxetable}{lcccc}
\tabletypesize{\scriptsize}
\tablecaption{Photometry with Fixed Angular Radius  Apertures
\label{photometry}}
\tablewidth{0pt}
\tablehead{
\colhead{UT Date}    & \colhead{$\Phi$\tablenotemark{a}}    & \colhead{V\tablenotemark{b}} &  \colhead{$H_{V}(C)$\tablenotemark{c}} & \colhead{$C_e [km^2]\tablenotemark{d}$} 
}

\startdata

September 28 & 0.2 &  22.64$\pm$0.01  & 17.76$\pm$0.01 & 2.2$\pm$0.2 \\
September 28 & 1.0 &  22.02$\pm$0.01  & 17.14$\pm$0.01 & 3.8$\pm$0.4  \\
September 28 & 4.0 &  21.17$\pm$0.03   & 16.29$\pm$0.03 & 8.4$\pm$0.8 \\
September 28 & 6.0 &  21.05$\pm$0.06   & 16.17$\pm$0.06 & 9.4$\pm$0.9 \\

October 08 & 0.2 &  22.75$\pm$0.01   & 17.75$\pm$0.01 & 2.2$\pm$0.2 \\
October 08 & 1.0 &  22.26$\pm$0.01   & 17.26$\pm$0.01 & 3.4$\pm$0.3 \\
October 08 & 4.0 &  21.86$\pm$0.03   & 16.86$\pm$0.03 & 5.0$\pm$0.5 \\
October 08 & 6.0 &  21.63$\pm$0.06   & 16.63$\pm$0.06 & 6.1$\pm$0.6 \\

December 07 & 0.2  & 22.84$\pm$0.01   &17.40$\pm$0.01 & 3.0$\pm$0.3 \\
December 07 & 1.0  & 22.13$\pm$0.01   & 16.69$\pm$0.01 & 5.8$\pm$0.6 \\
December 07 & 4.0  & 21.83$\pm$0.03   & 16.39$\pm$0.03 & 7.7$\pm$0.7 \\
December 07 & 6.0  & 21.87$\pm$0.06   & 16.43$\pm$0.06 & 7.4$\pm$0.7 \\

December 18 & 0.2  & 22.76$\pm$0.01  & 17.29$\pm$0.01 & 3.3$\pm$0.3 \\
December 18& 1.0  & 21.96$\pm$0.01   & 16.49$\pm$0.01 & 7.0$\pm$0.7 \\
December 18 & 4.0  & 21.30$\pm$0.03   & 15.86$\pm$0.03 & 12.5$\pm$1.3 \\
December 18 & 6.0  & 21.22$\pm$0.06   & 15.75$\pm$0.06 & 13.8$\pm$1.4 \\

\enddata


\tablenotetext{a}{Projected angular radius of photometry aperture, in arcseconds}
\tablenotetext{b}{Apparent V-band magnitude}
\tablenotetext{c}{Absolute magnitude computed assuming a C-type phase function c.f.~Equation (\ref{absolute})}
\tablenotetext{d}{Cross-section computed from $H_V(C)$ using Equation (\ref{area}) with $p_V$ = 0.05}

\end{deluxetable}

\clearpage 
\begin{deluxetable}{lcccccc}
\tabletypesize{\scriptsize}
\tablecaption{Photometry with Fixed Linear Radius Apertures
\label{absolute}}
\tablewidth{0pt}
\tablehead{
\colhead{UT Date}    & \colhead{Quantity\tablenotemark{a}} & \colhead{460 km }   & \colhead{2300 km} & \colhead{9200 km} & \colhead{13800 km}
}

\startdata

September 28 	&	V 			& 22.53$\pm$0.01	 	& 21.82$\pm$0.01	 	& 21.55$\pm$0.03	 		& --	 		 \\
September 28   & H$_V(C)$ 			& 17.65$\pm$0.01		& 16.94$\pm$0.01	 	& 16.67$\pm$0.03	 		& --	 		 \\
September 28 	& $C_e$ 			& 2.4$\pm$0.2 		& 4.6	$\pm$0.5	& 5.9$\pm$0.6			& --			 \\\\
	
October 08 	&	V 			& 22.64$\pm$0.01	& 22.21$\pm$0.01	 	& 21.93$\pm$0.03	 		& 21.99$\pm$0.06	 		 \\
October 08   & H$_V(C)$ 			& 17.64$\pm$0.01		& 17.21$\pm$0.01	 	& 16.93$\pm$0.03	 		& 16.99$\pm$0.06	 		 \\
October 08 	& $C_e$ 			& 2.4$\pm$0.2 		& 3.6$\pm$0.4		& 4.7$\pm$0.5			& 4.4$\pm$0.4			 \\\\

December 07 	&	V 			& 22.84$\pm$0.01	 	& 22.11$\pm$0.01	 	& 21.85$\pm$0.03	 		& 21.95$\pm$0.06	 		 \\
December 07   & H$_V(C)$ 			& 17.40$\pm$0.01		& 16.66$\pm$0.01	 	& 16.41$\pm$0.03	 		& 16.51$\pm$0.06	 		 \\
December 07 	& $C_e$ 			& 3.0$\pm$0.3 		& 6.0$\pm$0.6		& 7.5$\pm$0.8			& 6.9$\pm$0.7			 \\\\

December 18 	&	V 			& 22.76$\pm$0.01	 	& 21.96$\pm$0.01	 	& 21.30$\pm$0.03	 		& 21.22$\pm$0.06	 		 \\
December 18   & H$_V(C)$ 			& 17.29$\pm$0.01		& 16.49$\pm$0.01	 	& 15.83$\pm$0.03	 		& 15.75$\pm$0.06	 		 \\
December 18 	& $C_e$ 			& 3.3$\pm$0.3 		& 7.0$\pm$0.7		& 12.8$\pm$1.3			& 13.8$\pm$1.4			 \\\\

\enddata

\tablenotetext{a}{V = apparent V magnitude, $H_V$ = Absolute magnitude computed assuming a C-type phase function c.f.~Equation (\ref{absolute}), $C_e$ = effective scattering cross-section in km$^2$ computed from $H_V(C)$ using Equation (\ref{area}) with $p_V$ = 0.05}

\end{deluxetable}

%
%
%
%
%
%
%
%
%
%
%
%

\clearpage 
\begin{deluxetable}{lcccccc}
\tablecaption{Nucleus Photometry
\label{nucleus_phot}}
\tablewidth{0pt}
\tablehead{
\colhead{UT Date}    & \colhead{UT Start Time\tablenotemark{a}} & \colhead{$V$\tablenotemark{b}}   & \colhead{$H_V$\tablenotemark{c}} 
& \colhead{$C_e [km^2]$\tablenotemark{d}} 
}

\startdata

Sep 28 & 23:13 &	22.66	&	17.79 	& 	2.12 	\\
 & 	23:22 	&	22.62	&	17.75	& 	2.12 	\\
 & 	23:31 	&	22.63	&	17.76	& 	2.18 	\\
 & 	23:40 	&	--	&		--		& 	-- 	\\
 & 	23:57 	&	22.61	&	17.74	& 	2.22 		\\

Oct 08 & 09:09	&	22.77	&	17.77	&  	2.14 	\\
 & 	09:18	&	22.75	&	17.75	& 	2.18 	\\	
 & 	09:27	&	22.72	&	17.72	& 	2.24 		\\
 & 	09:37	&	22.70	&	17.70	& 	2.28 	\\
 & 	09:46	&	22.68	&	17.68	& 	2.33 	\\

Dec 07 & 03:41	&	22.85	&	17.42 	& 	2.98 	\\
 & 	03:50	&	22.82	&	17.38 	& 	3.07 	\\
 & 	03:59	&	22.83	&	17.39	& 	3.04 	\\
 & 	04:09	&	22.82	&	17.39	& 	3.07 	\\
 & 	04:18	&	--	&	--	& 	-- 	\\

Dec 18 & 00:32	&	22.78	&	17.32	& 	3.26\\
 & 	00:41	&	22.80	&	17.34	& 	3.20\\
 & 	00:50	&	22.75	&	17.29	& 	3.35\\
 & 	01:00	&	22.72	&	17.26	& 	3.45\\
 & 	01:09	&	22.71	&	17.25	& 	3.48\\
 & 	02:07	&	22.73	&	17.27	& 	3.41\\
 & 	02:16	&	22.74	&	17.28	& 	3.38\\
 & 	02:25	&	22.76	&	17.30	& 	3.32\\
 & 	02:35	&	22.76	&	17.30	& 	3.32\\
 & 	02:44	&	22.74	&	17.28	& 	3.38\\
 & 	03:42	&	--	&	--	& 	--			\\
 & 	03:51	&	--	&	--	& 	--			\\
 & 	04:00	&	22.76	&	17.30	& 	3.32\\
 & 	04:10	&	22.76	&	17.30	& 	3.32\\
 & 	04:19	&	22.75	&	17.29	& 	3.35\\

\enddata

\tablenotetext{a}{Start time of the observation.  }
\tablenotetext{b}{Apparent V magnitude within a 0.2\arcsec~radius photometry aperture.  Nominal photometric uncertainty $\pm$0.01 magnitudes.} 
\tablenotetext{c}{$H_V$ = Absolute magnitude computed assuming a C-type phase function c.f.~Equation (\ref{absolute}), }
\tablenotetext{d}{$C_e$ = effective scattering cross-section in km$^2$ computed from $H_V(C)$ using Equation (\ref{area}) with $p_V$ = 0.05}

\end{deluxetable}



\clearpage

\clearpage

\clearpage

\clearpage

\clearpage

\clearpage


\end{document}